\documentclass[10pt]{iopart}
\usepackage{graphicx}  
\usepackage{times}

\begin{document}
\renewcommand{\textheight}{7.48truein}

\title
[Clusters, phason elasticity, and entropic stabilisation]
{Clusters, phason elasticity, and entropic stabilisation:
a theory perspective}

\author{C.~L.~Henley\dag}

\address{\dag\ Dept. of Physics, Cornell University,
Ithaca NY 14853-2501}

\begin{abstract}
Personal comments are made about the title subjects, including:
the relation of Friedel oscillations to Hume-Rothery
stabilisation;
how calculations may resolve the random-tiling
versus ideal pictures of quasicrystals; and the role
of entropies apart from tile-configurational.
\end{abstract}

\section{Introduction} 

    My viewpoint comes from a bottom-up 
approach~\cite{Mihet02,Henet02} to modeling
quasicrystal structure and explaining their thermodynamic 
stabilisation.  That is, we start with ab-initio or pair
potential based evaluation of the total energy, to capture
the T=0 behaviour; or perhaps MD (Molecular Dynamics) and MC
(Monte Carlo) simulation for $T>0$.  We inductively identify
motifs and restrict our model, so it has freedom to explore
only the states we know are comparatively good. [At this 
level, the model may be formulated in terms of tilings
or clusters, but the discrete geometry is just standing
in as a way to label the distinct low-energy atomic configurations.]
New simulations are constructed, which can handle larger
length scales because there are fewer degrees of freedom.
At the moment, this appears to be the most 
direct approach to acertain whether energetic stabilisation
(e.g. an implementation of Penrose's matching rules) is
ever relevant to real quasicrystals.

\subsection{Importance of using realistic potentials}

\begin{figure}
\includegraphics[width=4.9in,angle=0]{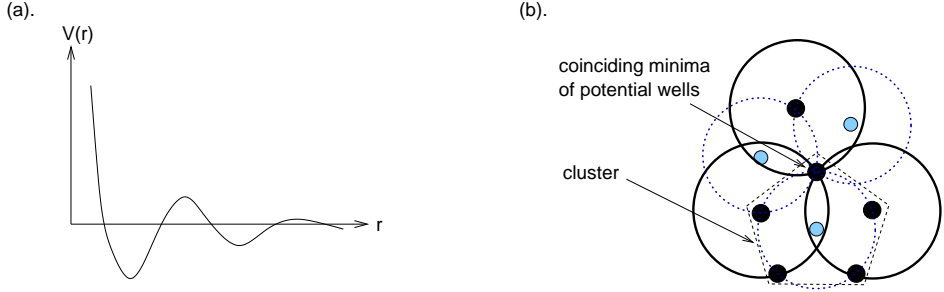}
\caption{
(a). Typical long-range potential with Friedel oscillations.
(b). Schematic example of cluster (dashed pentagon) stabilised from 
the outside.  The figure highlights interactions with the 
topmost atom in the cluster: it 
lies at the bottom of potential wells 
(shown by circles) from five atoms inside the cluster,
and from three more atoms outside it.}
\label{fig:potcluster}
\end{figure}

In a metal, realistic pair potentials have Friedel oscillations
[Fig.~\ref{fig:potcluster}(a)].
I will review why this is the real-space analog of the
`Hume-Rothery' (and related) mechanisms  in reciprocal space.
The energy from second-order perturbation theory,
computed in reciprocal space, is
     \begin{equation}
         \delta E^{(2)} =  - \frac{1}{2} \sum _{\bf k} \sum _{\bf q}
                    \frac {2 |\tilde{\phi}({\bf q})|^2  
                                 |\tilde \rho({\bf q})|^2}
                     {E({\bf k}+{\bf q}) - E({\bf k}) } , 
     \label{eq:E2}
     \end{equation}
where  $E({\bf k})$ is the free electron dispersion,
and $\phi({\bf r})$ is the [short ranged] potential
for an atom to scatter an electron.
[The factor of 2 is from electron spin;
I do not encumber this schematic formula with 
multiple atomic species.]
So $\delta E$ is most negative when 
$\tilde{\phi}({\bf G})$, the atom density at a reciprocal lattice
vector $\bf G$, is strong for $|{\bf G}|\approx 2 k_F$ 
(near to spanning the Fermi sphere):
that was Hume-Rothery's criterion.
If we now do the sum over ${\bf k}$ in (\ref{eq:E2}), 
lumping the prefactors and calling it
`${\tilde{V}_{\rm eff}}({\bf q})$',  we get
   \begin{equation}
     \delta E ^{(2)}
    = - \frac {1}{2} \sum _{\bf q} {\tilde{V}_{\rm eff}}({\bf q})
                  |{\tilde \rho}({\bf q})|^2
    \equiv  \frac{1}{2} \sum _{ij} V_{\rm eff}({\bf r}_i-{\bf r}_j),
   \end{equation}
where $\{ {\bf r}_i \}$ are the atoms' positions;
the second equality just comes from Fourier transforming, 
[Recall $\tilde \rho({\bf q}) =\sum _i e^{i {\bf q}\cdot{\bf r}_i}$.]
The effective interatomic potential $V_{\rm eff}(r)$ turns out to be
the electron susceptibility $\chi(r)$, 
convolved twice with $\phi(r)$ in three dimensions. 
Now the sharp Fermi surface in reciprocal space induces
Friedel oscillations in real space ---
a factor $\propto \cos(2 k_F r+\delta)$ in $\chi(r)$ ---
which are inherited by $V_{\rm eff}(r)$.    Thus, Hume-Rothery
stabilisation is practically equivalent to saying the second minimum
in $V_{\rm eff}(r)$ is essential for determining the structure.

At least, Friedel oscillations
are important in the Al-TM 
family [i(AlPdMn), d(AlNiCo)] and in the Frank-Kasper family [i(ZnMgRE)].  
In this conference, Mihalkovi\v{c} and Widom~\cite{widom-icq9}
assert that the
embedded-atom potentials
[which lack Friedel oscillations, but implicitly include multi-atom 
interactions] work excellently for the i(CdCa) family.

       As a result, the sites within a `cluster' are not governed 
by interactions with the other atoms in the cluster, but are a
resultant of overlapping spheres representing the potential wells
of second and third neighbour atoms, including those outside the
cluster.  
This seemed to me the only explanation for the pseudo Mackay cluster 
in e.g. $i$-AlCuFe.
[The outer shells have icosahedral symmetry, but
the innermost one has a roughly 1/3 filling of 20 sites on 3-fold 
axes, where one geometrically would have expected an icosahedron].

\subsection{Temperature does matter}

       The Al atoms have rather weak interactions, and sometimes seem
to behave almost like a fluid moving around a framework of 
fixed and well-ordered TM atoms.
Simulations by Cockayne  and Widom~\cite{widom-Al}, 
Henley {\it et al}~\cite{Henet02},
and Hocker and G\"ahler~\cite{gaehler-Al}
found frequent hoppings of the Al atoms, so that one wonders whether 
to speak of well-defined sites.  The configurational entropy from this is
obviously huge.

Simulational approaches have scarcely scratched the surface
as far as addressing thermally excited disorder, but I would say
we know what to do.  The reason it has not been tried is that
a thorough ab-initio-based study of a good quasicrystal has been carried out
just a handful of times so far, and it is much simpler to do it at $T=0$.

\section {Clusters}

The simulation experiences have led me to believe
that in the known quasicrystals,
clusters do not have physical reality 
(in the sense of having significantly stronger
intra-cluster bonding).
Yet clusters (or tiles, which I consider
to be a closely related concept) are inescapable as a 
framework to organise our understanding of a structure.

A lesson we learned from more fundamental 
fields of 20th-century physics is that we all need
frameworks to organise our thinking, but we forget that
it is a coordinate system we impose to 
decribe phenomena, not a physical reality.
We should not be seduced by our tools and ascribe fundamental 
significance to these mental constructs.  Let us instead be open to 
`complementarity': that is, dissimilar descriptions
may secretly be equivalent.

\subsection{Clusters for analyzing structures as predicted from
energy calculations}
\label{sec:cluster-deca}

Our experiences with decagonal structure models~\cite{Mihet02,Gu-icq9}
offer abundant examples in which the {\it same}  structure may be
expressed in terms of cluster $C$ on network $N$, or of
cluster $C'$ on network $N'$.  
Perhaps cluster $C'$ is
related to $C$ by some sort of inflation; or perhaps $C'$ is
associated with the voids in the packing of $C$ and
vice versa.  (An example is the complementarity of
Bergman clusters and Mackay clusters in  icosahedral
quasicrystals.)

      In a great many models, the cluster
is a mathematical corollary of a set of inequalities 
which express the energy minimisation.  This appears
most baldly, perhaps, in abstract tiling models
where one demands to maximise the frequency
of occurrence of cluster $M$ 
\cite{Jeong94,Hen97,reichert}
-- one presumably would
say that cluster $M$ does have a physical content in this
model.   But in many cases, this optimisation forces the 
presence of a much larger supertiling, and 
much larger clusters $C$ appear at its vertices.
Thus, the appearance of the cluster is dependent on
things that happen far away: 
a slight change in e.g. stoichiometry, 
and the clusters which dominated the whole image
may dissolve in favour of some other motif, as a consequence
of delicately competing energies. 

A second way in which the cluster configuration depends on 
faraway atoms can be seen in Al-TM quasicrystals, which
are well described using pair potentials that have
a strong second potential well  due to Friedel
oscillations.  The fact that an atom occurs in 
a particular site within a cluster is then 
not mainly due to interactions with its neighbours
in the cluster, but is the resultant of
potentials from many more distant neighbours 
in the space surrounding the cluster, and which 
are sufficiently correlated that the potentials from
them add constructively (see Fig.~\ref{fig:potcluster}(b)).

In conclusion, my proposed operational definition of
clusters is statistical: `a pattern of atoms which is found in 
all examples of a given ensemble.'
An ensemble is implicit in anybody's
definition of a cluster: to locate the boundary around a
group of atoms, it must be possible to surround the
cluster in more than one way; one could
not talk of, e.g., the fcc lattice as being built
from clusters. 
The implicit ensemble 
might be either multiple occurrences
of the cluster in a large unit cell, or
various crystal phases that contain the same
cluster, or a single simulation cell in which
one enumerates all the low-energy structures.

\subsection{Clusters in cleavage and interfaces}

One cannot rule out a cluster description a priori
as a way to map the energy landscape of a structure.
When I was more naive, in fact, I advanced a cluster
model which assumed a certain cost of cutting the 
linkages between Mackay clusters along 2-fold and
3-fold symmetry directions, so as to predict the
equilibrium crystal shape~\cite{Lei91}.  
Although a certain knobbliness
might be anticipated, still even
if the crevices got filled by Al atoms the
energy cost might still be a linear function of the
cluster-cluster linkages cut by the placement
of the interface.

But in view of the experience mentioned
in Sec.~\ref{sec:cluster-deca}, I no longer expect that
cluster linkages govern the energy differences due
to  offsets or orientations of the interface.  
Nevertheless, one can imagine `clusters' emerging
in the purely statistical sense expressed in the
`operational definition' at the end of Sec.~\ref{sec:cluster-deca}.
If a crack approaches similar groupings of atoms,
presumably it tends to pass through them in
similar fashions. Then the atoms which always find
themselves together on the same side of the crack
could be designated a `cluster' for the
purposes of describing crack propagation.
Notice that (i) such `clusters' need not
be the same ones that are useful in describing the
equilibrium ensemble and (ii) one would expect
their shape to have less symmetry than the
material itself, as it must depend on the 
orientations of the crack propagation and of the crack shear.
If it happens that in fact the {\it same} clusters appear in
cracks of various orientations, and for other physically
defined ensembles, that would be justification to attribute
a `physical reality' to the cluster -- but such tests seem
possible mainly in simulations, not in experiments.

\section{Phason elasticity}

My motivation here is not in the physical consequences of phason 
elasticity, but in using elasticity as an indicator of the nature
of the quasicrystal state. 
{\it Gradient-squared}
elastic free energies appear only in the {\it random-tiling} kind of phase, 
in the sense distinguished in Sec.~\ref{sec:stabilisation}, above.  
Matching-rule interactions
would lead to an energy cost which is {\it linear} in the 
absolute value of phason strain components~\cite{socolar86},
presumably the same is true for any other interaction
that has the same ground state (or to one in the same
`local isomorphism' class). 

A small caveat should be offered.  
The linear cost is related to the discrete hops that 
are mathematically unavoidable in structures with the
usual quasicrystal space groups~\cite{frenkel86}.
On the other hand, for unusual non-symmorphic space groups,
a `continuous phason mode' is possible~\cite{levitov89a,levitov89b}
which {\it may} exhibit gradient-squared elasticity in 
the ground state, in the same sense that one-dimensional 
incommensurate crystals may.  But no plausible atomic
model structure of this class has ever been exhibited,
much less a set of interactions for which it is the
ground state.

The experiments of de Boissieu 
{\it et al}~\cite{deboissieu95,francoual05}
on icosahedral phases are the only ones I know that support 
the validity of elastic theory and thus, implicitly, a 
random-tiling-like equilibrium state.
But I will not fully trust the elastic interpretation of 
diffuse scattering till the quantitative elastic constants 
agree (at least in order of magnitude) with a plausible simulation.
That has not happened yet: one reason is that the
`canonical cell' tiling, which is the simplest way
to make a well-specific ensemble for most cluster-based
icosahedral models,~\cite{widom-icq9,Mihet96}
is also the least tractable tiling to simulate~\cite{mih04};
a second reason
is that it is nontrivial to extract an absolute scale
of fluctuations from measurements  of diffuse scattering.

In the case of {\it decagonals,}
no evidence of gradient-squared elasticity has ever been seen
in experiments. The behaviour of the entropic elastic theory
in decagonal random tiling models is not well understood, either.
Thus, it seems more plausible to me that 
matching rules (or the equivalent) are realised in 
decagonal quasicrystals, than  in icosahedral ones.

\section{Thermodynamic stabilisation of quasicrystal phases}
\label{sec:stabilisation}

The two fundamental competing scenarios of the
stabilisation of quasicrystals 
are not exactly `entropy' versus `energy', as we often
loosely say.   Rather, the question is whether 
the  model is in the qualitative class that contains
quasiperiodic ideal tilings that have purely Bragg peaks,
or the class that contains the maximally random tiling 
in which long range order is an emergent phenomenon.
This distinction is a rigorous one from the viewpoint
of statistical mechanics, because these two states
are separated by a phase transition, but that is of
no help in distinguishing them experimentally.

\subsection{Role of simulations}

I have come to believe that ab initio modeling,
though tedious on account of the many levels of
description between 
microscopic and macroscopic~\cite{Mihet02,Henet02},
is the quickest 
path to a solid understanding of which scenario should
apply to a particular given material.
It is sometimes objected that simulation is unfeasible
for handling e.g. incommensurate modulations, when the effective
repeat cell is far too large to simulate by brute force.
But there are generally ways to bridge to large scales by
connecting the simulations first to a kind of continuum model.
(A valid analogy is that one can understand the geometry of a large
soap bubble by evaluating its surface tension, which can be 
computed using a far smaller simulation cell.)

To get meaningful results, it is crucial (and very
difficult) that the structure used in the modeling be made
consistent with the Hamiltonian assumed -- 
it must be the ground state (in the $T=0$ case), or nearly so. 

Rather generally, 
Al-TM quasicrystals seem to have a framework
of well-fixed sites plus a scattering of sensitive sites.
In $i$-AlMnSi (and perhaps some others), these are
the $\delta$ atoms of Ref.~\cite{Mihet96}, located on 6D
body centres in the hyperspace formulation, or centring
Bergman clusters in real space, which can alternate
between Al or TM atoms or vacancies.  In $d$-AlCoNi
(Ni-rich or Co-rich), there are Al atoms in `channels'~\cite{Gu-icq9}
which undergo (in simulated models) intricate occupational/displacive
orderings.   
At a higher level of description, such atomic orderings can be expressed 
as terms -- maybe the dominant terms -- in the effective tile-tile 
interaction (whether or not it realises matching rules).

\subsection{Stoichiometry in random-tiling models?}

It was asserted that `a considerable amount of chemical disorder
is essential to a random-packing model.~\cite{e-abe}'.
Perhaps this is based on a mistaken picture that
random packings (of clusters) necessarily create 
atom conflicts that must be resolved in a context-dependent way.
But actually each {\it tile} in a tiling
-- whether random or governed by local rules --
has a finite set of local environments, and the
atom decoration is {\it designed} to fit well with every
environment, without overlapping or conflicting atoms.

In the simplest cases -- e.g. the square-triangle tiling with 
dodecagonal symmetry, or the rhombohedral tiling with ico
symmetry -- there are two kinds of tiles, and the (irrational)
ratio of their numbers is fixed by the symmetry.  The 
decoration is deterministic, so there is a unique stoichiometry.
However, tilings related to real quasicrystals often have
more than two tiles --
e.g. the `canonical cell tiling'~\cite{widom-icq9,Mihet96}
with ico symmetry,
or the `hexagon-boat-star' tiling~\cite{Mihet02,widom-Al} 
with decagonal symmetry.
Then within the sum rules fixed by symmetry, one kind of tile can be
traded for a combination of the others; sometimes the decoration
is such that this trade leaves the atom count unchanged~\cite{Mihet02},
so our model is line compound in this case too. In the case that
the atom content changed, though, the alloy composition would indeed
be variable; but as Elser once suggested,
such a decoration is undesirable since phason dynamics
would be coupled to, and slowed by, mass diffusion, 
contrary to the observed relaxation in good quasicrystals.

\subsection{Origin of matching rules?}

Note that although it has become attractive to reformulate matching 
rules in terms of a decoration that implements Gummelt-like
covering rules~\cite{e-abe}, I do not think this is likely to emerge
from an atomistic model.  If a covering cluster emerges,
it will be a sort of accident from the maximimation of
some smaller non-overlapping cluster (see my comments
on clusters in Sec.~\ref{sec:cluster-deca}).

Instead, I consider it much likelier that something emerges
similar in spirit to Penrose's arrow rules.  That is: the
larger energy scales expressed by the fundamental 
atomic sites define a random-tiling ensemble; 
then additional occupational orderings in relatively
rare sites define (something like) a set of interacting
arrows.  In fact, Widom~\cite{widom-AlCuCo-1,widom-AlCuCo-2}  
and coworkers have found (in the past few years) a matching rule
{\it almost} equivalent to Penrose's, in the
decagonal $d$(AlCuCo), implemented in that case
by an alternation AlCo/CoAl in the chemical occupancy
of a pair.  

\subsection{Stabilisation by which entropy?}

The evidence is abundant that many quasicrystals are
high-temperature phases, ergo stabilised  by entropy -- 
but, in many cases, not the tiling configurational entropy,
for that is too small.
So, the larger entropy of vibrations or of chemical disorder is the
only candidate to affect the phase diagram.
In the past, we brushed aside
such entropy contributions, by claiming they must have a similar
value in the quasicrystal phase as they do in the approximant phases.
The latter phases were assumed
to be the immediate competitors of the quasicrystal phase
in the phase diagram, so the difference  in vibrational
or substitutional free energies would largely cancel.
But perhaps
we need to examine more carefully just how the quasicrystal
ordering may affect these entropic terms.

\ack
I am grateful to
Marc de~Boissieu,
Veit Elser,
Marek Mihalkovi\v{c},
Walter Steurer,
and Mike Widom,
for conversations and communications.

This work was supported by 
the U.S. Department of Energy grant
DE-FG02-89ER-45405.

\end{document}